\documentclass[12pt, a4paper] {article}
\usepackage{amsfonts}
\usepackage{verbatim}
\usepackage{color}

\begin{document}

\begin{center}
{\sf\Large{Abelian 3-form gauge theory: superfield approach}} 

\vskip 2.5cm

{\sf{\small R. P. MALIK}}\footnote{Talk delivered at BLTP, JINR, Dubna, Moscow Region, Russia
in the International Workshop on ``Supersymmetries and Quantum Symmetries'' (SQS'11) [18-23 July 2011].
Slightly modified versions of this article were presented at International Conferences: QFT-11, IISER, 
Pune [23-27 February 2011] and
NTFT-2, BHU, Varanasi [7-12 February 2011].} \\
{\it Department of Physics, Centre of Advanced Studies,}\\
{\it Banaras Hindu University, Varanasi-221 005, (U. P.), India}\\

\vskip 0.1cm

{\sf and}\\

\vskip 0.1cm

{\it DST Centre for Interdisciplinary Mathematical Sciences,}\\
{\it Banaras Hindu University, Varanasi-221 005, (U. P.), India}\\
{\small e-mails: rudra.prakash@hotmail.com, malik@bhu.ac.in}   

\vskip 2cm

\end{center}

\noindent
{\bf Abstract:} We discuss a $D$-dimensional Abelian 3-form gauge theory within the framework
of Bonora-Tonin's superfield formalism and derive the off-shell nilpotent and 
absolutely anticommuting Becchi-Rouet-Stora-Tyutin (BRST) and anti-BRST symmetry transformations
for this theory. To pay our homage to Victor I. Ogievetsky (1928-1996), who was one of the inventors
of Abelian 2-form (antisymmetric tensor) gauge field, we go a step further and discuss the above $D$-dimensional 
Abelian 3-form gauge theory
within the framework of BRST formalism and establish that the existence of the (anti-)BRST invariant
Curci-Ferrari (CF) type of restrictions is the hallmark of any arbitrary $p$-form gauge theory
(discussed within the framework of BRST formalism).\\



\newpage

\section{Introduction}

In recent years, the study of higher $p$-form ($p = 2, 3, 4,...$) gauge theories has
become quite fashionable because of its relevance in the context of (super)string theories and
related extended objects (see, e.g. [1,2]). It is worthwhile to mention, in this context,
that Ogievetsky and Polubarinov [3,4] were the first to study the 2-form antisymmetric
tensor gauge field (way back in 1966-67). Our presentation is a tribute to V. I. Ogievetsky
(1928-96) because we go a step further in the direction of the study of higher $p$-form
($p \geq 2$) gauge theories and discuss the Abelian 3-form gauge theory in arbitrary
$D$-dimensions of spacetime within the framework of superfield approach to Becchi-Rouet-Stora-Tyutin
(BRST) formalism, proposed in [5,6].

We derive the proper (i.e. off-shell nilpotent and absolutely anticommuting)
(anti-)BRST symmetry transformations for the above 3-form [$A^{(3)} = (1/3!) 
(dx^\mu \wedge dx^\nu \wedge dx^\eta) A_{\mu\nu\eta}$] totally antisymmetric 
tensor gauge field $A_{\mu\nu\eta}$ which appears in the quantum excitations 
of the (super)strings. Furthermore, we obtain the proper (anti-)BRST transformations
associated with the (anti-)ghost fields of the theory. Our main goal is to establish that the existence of the
Curci-Ferrari (CF) type restrictions [7] is the hallmark of any arbitrary $p$-form
($p = 1, 2, 3,...$) gauge theory when it is discussed within the framework of 
superfield approach to BRST formalism. In fact, we show that the derivation of the
CF-type condition(s) is a very natural consequence of the application
of the superfield approach [5,6] to BRST formalism.

Our present write-up is organized as follows. In Sec. 2, we recapitulate the bare
essentials of the horizontality condition and apply it to the Abelian 3-form gauge
theory. In the next section, we derive the proper (anti-)BRST symmetry transformations
and corresponding coupled (but equivalent) Lagrangian densities
for the present theory. Our Sec. 4 deals with the critical and crucial comments on the CF type
restrictions. Finally, in Sec. 5, we summarize our key results and make some concluding remarks.

\section{Horizontality Condition}

Let us begin with the starting Lagrangian density of the Abelian 3-form field
\begin{eqnarray}
{\cal L}_0 = \frac{1}{24} H^{\mu\nu\eta\kappa} H_{\mu\nu\eta\kappa}, \quad
H_{\mu\nu\eta\kappa} = \partial_\mu A_{\nu\eta\kappa} - \partial_\nu A_{\eta\kappa\mu}
+ \partial_\eta A_{\kappa\mu\nu} - \partial_\kappa A_{\mu\nu\eta},
\end{eqnarray}
where the 4-form $H^{(4)} = d A^{(3)} \equiv (1/4!) (dx^\mu \wedge dx^\nu \wedge dx^\eta \wedge dx^\kappa) H_{\mu\nu\eta\kappa}$ defines the curvature tensor $H_{\mu\nu\eta\kappa}$ which is derived from the exterior derivative
$d = dx^\mu \partial_\mu$ ($d^2 = 0, \mu, \nu, \eta,...= 0, 1, 2, ..., D-1$) and the 3-form $A^{(3)}$
that encodes the totally antisymmetric tensor gauge connection $A_{\mu\nu\eta}$. It can be easily seen that, under the following local gauge symmetry transformations $\delta_g$ (with the local infinitesimal 
antisymmetric gauge parameter $\Lambda_{\mu\nu} = - \Lambda_{\nu\mu}$)
\begin{eqnarray}
\delta_g A_{\mu\nu\eta} = \partial_\mu \Lambda_{\nu\eta} + \partial_\nu \Lambda_{\eta\mu} + \partial_\eta
\Lambda_{\mu\nu},
\end{eqnarray}
the Lagrangian density and the curvature tensor $H_{\mu\nu\eta\kappa}$ remain invariant.
Thus, the curvature tensor $H_{\mu\nu\eta\kappa}$, derived from the 4-form $H^{(4)}$, is a physical 
quantity (in some sense). One of the most intuitive approaches to quantize the $p$-form gauge theory 
is the BRST formalism where the local gauge symmetry transformations (e.g. (2)) are traded with 
the proper (i.e. nilpotent and absolutely anticommuting) (anti-)BRST symmetry transformations. 

The above (anti-)BRST symmetries can be derived by exploiting the geometrical superfield formalism [5,6]
where any arbitrary $D$-dimensional gauge theory is generalized onto the $(D, 2)$-dimensional 
supermanifold as [8]:
\begin{eqnarray}
&& d \longrightarrow \tilde d \equiv dZ^M \partial_M = dx^\mu \;\partial_\mu 
+ d \theta \;\partial_\theta + d {\bar \theta}\; 
\partial_{\bar \theta}, \quad x^\mu \longrightarrow Z^M = (x^\mu, \theta, \bar \theta), \nonumber\\
&& A^{(3)} \longrightarrow \tilde A^{(3)} = \frac {(dZ^M \wedge dZ^N \wedge dZ^K)}{3!}\; \tilde A_{MNK},
\qquad \partial_M = (\partial_\mu, \partial_\theta, \partial_{\bar \theta}).
\end{eqnarray}
Here the superspace variables $Z^M = (x^\mu, \theta, \bar \theta)$ are the generalization 
of the ordinary $D$-dimensional coordinates $x^\mu$ that incorporate a pair of Grassmannian variables 
$\theta$ and $\bar \theta$, too, which satisfy $\theta^2 = \bar \theta^2 = 0, \theta \bar\theta + 
\bar \theta \theta = 0$. In the horizontality condition (HC), we require the physical (geometrical) quantity 
$H^{(4)}$ to remain independent of the Grassmannian variables, namely;
\begin{eqnarray}
\tilde H^{(4)} = H^{(4)}\;\; \Longrightarrow \;\;\tilde d \tilde A^{(3)} = d A^{(3)}.
\end{eqnarray} 
This leads, automatically, to the derivation of the off-shell nilpotent $(s_{(a)b}^2 = 0)$ and 
absolutely anticommuting $(s_b s_{ab} + s_{ab} s_b = 0 )$ (anti-)BRST symmetry transformations $(s_{(a)b})$
for the gauge field and corresponding (anti-)ghost fields of the theory which we discuss, in detail, 
in the following section.

\section{Proper (Anti-)BRST Symmetries} 

It is clear from equation (4) that the r.h.s. of it contains only the spacetime differentials 
[i.e. $d A^{(3)} = (1/4!)(dx^\mu \wedge dx^\nu \wedge dx^\eta \wedge dx^\kappa)H_{\mu\nu\eta\kappa}$].
However, the l.h.s. contains the spacetime differentials together with the Grassmann 
differentials. Thus, it is evident that the HC amounts to setting equal to zero the coefficient of
all the differentials that contain Grassmannian variables. To check the above statement, it is 
imperative to compute the l.h.s. explicitly. Towards this goal, it can be seen that the equation 
(3) implies [8]
\begin{eqnarray}
&& \tilde A^{(3)} = {\displaystyle \frac{1}{3!}\; (dx^\mu \wedge dx^\nu \wedge dx^\eta)\;
\tilde A_{\mu\nu\eta} \;+ \;
\frac{1}{2} (dx^\mu \wedge dx^\nu \wedge d\theta) \tilde
A_{\mu\nu\theta}} \nonumber\\ && + {\displaystyle \frac{1}{2} (dx^\mu \wedge dx^\nu \wedge d \bar\theta)
\tilde A_{\mu\nu\bar\theta} + \frac{1}{3!} (d\theta \wedge d\theta
\wedge d\theta) \tilde A_{\theta\theta\theta} +  \frac{1}{3!} (d
\bar\theta \wedge d\bar\theta \wedge d \bar\theta) \tilde A_{\bar\theta\bar\theta\bar\theta}}
\nonumber\\ && + {\displaystyle (dx^\mu \wedge d \theta \wedge d\bar\theta) \tilde A_{\mu\theta\bar\theta} + \frac{1}{2}
(dx^\mu \wedge d\theta \wedge d \theta) \tilde A_{\mu\theta\theta}
+ \frac{1}{2} (d\theta \wedge d\theta \wedge d \bar\theta) \;\tilde A_{\theta\theta\bar\theta} }
 \nonumber\\
&& + {\displaystyle \frac{1}{2} (dx^\mu \wedge d\bar\theta \wedge d \bar\theta) \tilde A_{\mu\bar\theta\bar\theta}
+ \frac{1}{2} (d\theta \wedge d\bar\theta \wedge d \bar\theta) \;\tilde A_{\theta\bar\theta\bar\theta} }.
\end{eqnarray}
Keeping in mind the nature of the superfields, we make the suitable identifications: 
$\tilde A_{\mu\nu\eta} = \tilde {\cal A}_{\mu\nu\eta} (x, \theta, \bar \theta)$,
$\tilde A_{\mu\nu\theta} = \tilde {\bar {\cal F}}_{\mu\nu} (x, \theta, \bar\theta)$, 
$\tilde A_{\mu\nu\bar\theta} = \tilde {\cal F}_{\mu\nu} (x,\theta,\bar\theta)$,
$\tilde A_{\mu\theta\bar\theta} = \tilde \Phi_\mu
(x,\theta,\bar\theta)$, $\frac{1}{3!} \tilde A_{\theta\theta \theta} = \tilde
{\bar {\cal F}}_2 (x,\theta,\bar\theta)$, $\frac{1}{3!}
A_{\bar\theta\bar\theta\bar\theta} = \tilde {\cal F}_2  (x,\theta,\bar\theta)$,
$\frac{1}{2} \tilde A_{\theta\bar\theta\bar\theta} = \tilde {\cal F}_1
(x,\theta,\bar\theta)$, $\frac{1}{2} \tilde A_{\theta\theta\bar\theta} = \tilde 
{\bar {\cal F}}_1 (x,\theta,\bar\theta)$, $\frac{1}{2} \tilde A_{\mu\bar\theta\bar\theta}
= \tilde \beta_\mu (x, \theta, \bar\theta)$ and $\frac{1}{2} \tilde A_{\mu\theta\theta}
= \tilde {\bar \beta}_\mu (x, \theta, \bar\theta)$ as the generalization of the $D$-dimensional local fields $A_{\mu\nu\eta}, \bar C_{\mu\nu}, C_{\mu\nu}, \phi_\mu,$
$\bar C_2, C_2, C_1, \bar C_1, \beta_\mu$, $\bar \beta_\mu $ of the (anti-)BRST invariant
local $D$-dimensional ordinary theory onto 
the $(D, 2)$-dimensional supermanifold (within our superfield formalism).

The super-expansions of the above superfields, along the Grassmannian directions of the $(D, 2)$-dimensional
supermanifold are as follows [8]:
\begin{eqnarray}
\tilde {\cal A}_{\mu\nu\eta} (x, \theta, \bar\theta) &=& A_{\mu\nu\eta}
(x) + \theta\; \bar R_{\mu\nu\eta} (x) + \bar\theta\; R_{\mu\nu\eta} (x) +
i \;\theta \; \bar\theta\; S_{\mu\nu\eta} (x), \nonumber\\ \tilde \beta_\mu
(x, \theta, \bar\theta ) &=& \beta_\mu (x) + \theta \;\bar f^{(1)}_\mu (x) +
\bar\theta\; f^{(1)}_\mu (x) + i\; \theta\; \bar\theta\; b_\mu (x),
\nonumber\\ \tilde {\bar \beta}_\mu (x, \theta, \bar\theta) &=& 
\bar\beta_\mu (x) + \theta \;\bar f^{(2)}_\mu (x) + \bar \theta\; f^{(2)}_\mu (x) 
+ i\;\theta\;\bar\theta\; \bar b_\mu (x), \nonumber\\ \tilde \Phi_\mu (x,
\theta, \bar\theta) &=& \phi_\mu (x) + \theta \;\bar f^{(3)}_\mu (x) +
\bar\theta\; f^{(3)}_\mu (x) + i \;\theta \;\bar\theta\; b^{(3)}_\mu (x),
\nonumber\\ \tilde {\cal F}_{\mu\nu} (x, \theta, \bar\theta) &=& C_{\mu\nu}
(x) + \theta \;\bar B^{(1)}_{\mu\nu} (x) + \bar\theta\; B^{(1)}_{\mu\nu} (x)
+ i \;\theta \; \bar\theta\; s_{\mu\nu} (x), \nonumber\\ \tilde
{\bar {\cal F}}_{\mu\nu} (x, \theta, \bar\theta) &=& \bar C_{\mu\nu} (x) +
\theta\; \bar B^{(2)}_{\mu\nu} (x) + \bar\theta\; B^{(2)}_{\mu\nu} (x) + i
\; \theta\; \bar\theta \;\bar s_{\mu\nu} (x), \nonumber\\
\tilde {\cal F}_1 (x, \theta, \bar\theta) &=& C_1 (x) + \theta\;\bar b_1^{(1)} (x)
+ \bar \theta\; b_1^{(1)} (x) + i\;\theta\;\bar\theta\; s_1 (x), \nonumber\\
\tilde {\bar {\cal F}}_1 (x, \theta, \bar\theta) &=& \bar C_1 (x) + \theta\;\bar b_1^{(2)} (x)
+ \bar \theta\; b_1^{(2)} (x) + i\;\theta\;\bar\theta\; \bar s_1 (x), \nonumber\\
\tilde {\cal F}_2 (x, \theta, \bar\theta) &=& C_2 (x) + \theta\;\bar b_2^{(1)} (x)
+ \bar \theta\; b_2^{(1)} (x) + i\;\theta\;\bar\theta\; s_2 (x), \nonumber\\
\tilde {\bar {\cal F}}_2 (x, \theta, \bar\theta) &=& \bar C_2 (x) + \theta\;\bar b_2^{(2)} (x)
+ \bar \theta\; b_2^{(2)} (x) + i\;\theta\;\bar\theta\; \bar s_2 (x), 
\end{eqnarray}
where $A_{\mu\nu\eta}$ is the gauge field, $\phi_\mu$ is the vector bosonic field, 
$(\bar C_{\mu\nu})C_{\mu\nu}$ are 
the pair of fermionic antisymmetric (anti-)ghost fields, $(\bar \beta_\mu)\beta_\mu$ are the bosonic
ghost-for-ghost (anti-)ghost fields, $(\bar C_2)C_2$ and $(\bar C_1)C_1$ are the Lorentz scalar fermionic 
(anti-)ghost fields. The above fields are required for the proof of unitarity in the theory. The rest
of the fields, on the r.h.s. of equation (6), are secondary fields that have to be determined in terms
of the basic and auxiliary fields of the $D$-dimensional ordinary theory by exploiting the HC.

Explicit computation of (4) and setting equal to zero all the coefficients of the Grassmannian
differentials of the super 4-form of the l.h.s., leads to
\begin{eqnarray}
&& b_2^{(1)} = 0, \quad s_2 = 0, \quad \bar b_2^{(2)} = 0, \quad \bar s_2 = 0, \quad
\bar s_1 = 0, \quad s_1 = 0,  \nonumber\\
&& \bar b_2^{(1)} + b_1^{(1)} = 0, \quad b_1^{(2)} + \bar b_1^{(1)} = 0, \quad \bar f_\mu^{(2)}
= \partial_\mu \bar C_2, \quad f_\mu^{(1)} = \partial_\mu C_2, \nonumber\\
&& \bar b_\mu = - i \partial_\mu b_2^{(2)}, \quad b_\mu^{(3)} = - i \partial_\mu b_1^{(2)},
\quad B^{(1)}_{\mu\nu} = \partial_\mu \beta_\nu - \partial_\nu \beta_\mu, \quad
\bar f_\mu^{(2)} = \partial_\mu \bar C_2, \nonumber\\
&& \bar B_{\mu\nu}^{(2)} = \partial_\mu \bar \beta_\nu - \partial_\nu \bar \beta_\mu, \quad
s_{\mu\nu} = i (\partial_\mu \bar f_\nu^{(1)} - \partial_\nu \bar f_\mu^{(1)}) \equiv
- i (\partial_\mu  f_\nu^{(3)} - \partial_\nu  f_\mu^{(3)}), \nonumber\\
&& \bar s_{\mu\nu} = + i (\partial_\mu \bar f_\nu^{(3)} - \partial_\nu \bar f_\mu^{(3)}) \equiv
- i (\partial_\mu  f_\nu^{(2)} - \partial_\nu  f_\mu^{(2)}), \quad  b_\mu = i \partial_\mu \bar b_2^{(1)}, \nonumber\\ && R_{\mu\nu\eta}
= \partial_\mu C_{\nu\eta} + \partial_\nu C_{\eta\mu} + \partial_\eta C_{\mu\nu}, \quad
 \bar R_{\mu\nu\eta}
= \partial_\mu \bar C_{\nu\eta} + \partial_\nu \bar C_{\eta\mu} + \partial_\eta \bar C_{\mu\nu},
\nonumber\\&&  S_{\mu\nu\eta} = - i (\partial_\mu B_{\nu\eta}^{(2)} + \partial_\nu B_{\eta\mu}^{(2)}
+ \partial_\eta B_{\mu\nu}^{(2)}) \nonumber\\
&&\equiv
 + i (\partial_\mu \bar B_{\nu\eta}^{(1)} + \partial_\nu \bar B_{\eta\mu}^{(1)}
+ \partial_\eta \bar B_{\mu\nu}^{(1)}), \qquad b_2^{(2)} + \bar b_1^{(2)} = 0. 
\end{eqnarray}
In addition to the above results, we obtain the following Curci-Ferrari type restrictions
from the HC illustrated in (4), namely;
\begin{eqnarray}
f_\mu^{(2)} + \bar f_\mu^{(3)} = \partial_\mu \bar C_1, \quad
 \bar f_\mu^{(1)} +  f_\mu^{(3)} = \partial_\mu  C_1, \quad
\bar B_{\mu\nu}^{(1)} + B_{\mu\nu}^{(2)} = \partial_\mu \phi_\nu - \partial_\nu \phi_\mu,
\end{eqnarray}
which ensure the consistency of the {\it three} equivalences shown in (7). At this stage, a couple of remarks are
in order. First, the above restrictions emerge from setting the specific coefficients of the 4-form differentials
[e.g. $(dx^\mu \wedge d\theta \wedge d\theta \wedge d \bar\theta),
(dx^\mu \wedge d\theta \wedge d \bar \theta \wedge d \bar\theta), 
(dx^\mu \wedge d x^\nu \wedge d\theta \wedge d \bar\theta)$] of the l.h.s. of the HC.  Second, it is worth pointing out that the coefficients of
the differentials $(dx^\mu \wedge dx^\nu \wedge dx^\eta \wedge dx^\xi)$ from the l.h.s. and r.h.s.
of the condition $\tilde d \tilde A^{(3)} = d A^{(3)}$ match due to the precise form
of $R_{\mu\nu\eta}, \bar R_{\mu\nu\eta} $ and $S_{\mu\nu\eta}$,  quoted in (7).

To make the notations cute and a bit simpler, we identify:
$b_1^{(2)} = B_1, b_2^{(2)} = B_2, \bar b_2^{(1)} = \bar B,  \bar f_\mu^{(1)} =  F_\mu, f_\mu^{(2)} = \bar F_\mu,
f_\mu^{(3)} = f_\mu, \bar f_\mu^{(3)} = \bar f_\mu, B_{\mu\nu}^{(2)} = B_{\mu\nu}, 
\bar B_{\mu\nu}^{(1)} = \bar B_{\mu\nu}$.  As a consequence, the celebrated CF type restrictions become
$ B_{\mu\nu} + \bar B_{\mu\nu} = \partial_\mu \phi_\nu - \partial_\nu \phi_\mu,
f_\mu +  F_\mu = \partial_\mu C_1,
\bar f_\mu + \bar F_\mu = \partial_\mu \bar C_1 $. Furhermore, the {\it proper}
off-shell nilpotent and absolutely anticommuting (anti-)BRST symmetry transformations
that emerge from HC [cf. equation (4)] are [8,9]
\begin{eqnarray}
&& s_{ab} A_{\mu\nu\eta} = \partial_\mu \bar C_{\nu\eta} 
+ \partial_\nu \bar C_{\eta\mu} + \partial_\eta \bar C_{\mu\nu},
\qquad s_{ab} \bar C_{\mu\nu} = \partial_\mu \bar \beta_\nu - \partial_\nu \bar\beta_\mu, 
 \nonumber\\
&& s_{ab} \bar \beta_\mu = \partial_\mu \bar C_2, \qquad s_{ab} \bar C_2 = 0, \qquad 
s_{ab} \bar B_{\mu\nu} = 0, \qquad
s_{ab} C_1 = -  B_1,  \nonumber\\
&& s_{ab} \bar C_1 = - B_2, \qquad s_{ab} \bar B = 0, \qquad s_{ab} C_2 = \bar B, 
\quad s_{ab}  \beta_\mu = \bar F_\mu,
\quad s_{ab} \bar F_\mu = 0,  \nonumber\\
&& s_{ab} \bar f_\mu = 0, \quad s_{ab}  F_\mu = - \partial_\mu B_2, \quad 
s_{ab}  f_\mu = - \partial_\mu B_1,
\quad s_{ab} B_{\mu\nu} = \partial_\mu \bar f_\nu - \partial_\nu \bar f_\mu, \nonumber\\
&& s_{ab}  C_{\mu\nu} = \bar B_{\mu\nu}, \qquad  s_{ab} B_1 = 0, \qquad s_{ab}  B_2 = 0,
\qquad s_{ab} \phi_\mu = \bar f_\mu,
\end{eqnarray}
\begin{eqnarray}
&& s_b A_{\mu\nu\eta} = \partial_\mu C_{\nu\eta} + \partial_\nu C_{\eta\mu} + \partial_\eta C_{\mu\nu},
\qquad s_b C_{\mu\nu} = \partial_\mu \beta_\nu - \partial_\nu \beta_\mu,  \nonumber\\
&& s_b \beta_\mu = \partial_\mu C_2, \quad s_b C_2 = 0, \quad s_b B_{\mu\nu} = 0, \quad
s_b C_1 = - \bar B,  \nonumber\\
&& s_b \bar C_1 = B_1, \quad s_b B_1 = 0, \quad s_b \bar C_2 = B_2, \quad s_b \bar \beta_\mu = F_\mu,
\quad s_b F_\mu = 0, \nonumber\\
&& s_b f_\mu = 0, \quad s_b \bar F_\mu = - \partial_\mu \bar B, \quad 
s_b \bar f_\mu = \partial_\mu B_1,
\quad s_b \bar B_{\mu\nu} = \partial_\mu f_\nu - \partial_\nu f_\mu, \nonumber\\
&& s_b \bar C_{\mu\nu} = B_{\mu\nu}, \qquad  s_b \bar B = 0, \qquad s_b  B_2 = 0, \qquad s_b \phi_\mu = f_\mu.
\end{eqnarray}
It is elementary to check that the above 
(anti-)BRST symmetry transformations are off-shell nilpotent of order two
(i.e. $s_{(a)b}^2 = 0$).

The above transformations have been obtained from the superfield formalism without any knowledge of
the (anti-)BRST invariant Lagrangian density. This is due to the fact that
the substitution of the results of (7) into (6) leads to the following super-expansion
of the superfields in the language of the nilpotent
(anti-)BRST symmetry transformations [8]
\begin{eqnarray}
\tilde {\cal A}^{(h)}_{\mu\nu\eta} (x, \theta, \bar\theta) &=& A_{\mu\nu\eta}
(x) + \theta (s_{ab} A_{\mu\nu\eta} (x)) + \bar\theta (s_b A_{\mu\nu\eta} (x)) +
 \theta  \bar\theta (s_b s_{ab} A_{\mu\nu\eta} (x)), \nonumber\\ 
\tilde \beta^{(h)}_\mu
(x, \theta, \bar\theta ) &=& \beta_\mu (x) + \theta \;(s_{ab} \beta_\mu (x)) +
\bar\theta\; (s_b \beta_\mu (x)) +  \theta\; \bar\theta\; (s_b s_{ab} \beta_\mu (x)),
\nonumber\\ \tilde {\bar \beta}^{(h)}_\mu (x, \theta, \bar\theta) &=& 
\bar\beta_\mu (x) + \theta \;(s_{ab} \bar\beta_\mu (x)) 
+ \bar \theta\; (s_b \bar\beta_\mu (x)) 
+ \theta\;\bar\theta\; (s_b s_{ab} \bar\beta_\mu (x)), \nonumber\\ 
\tilde \Phi^{(h)}_\mu (x,
\theta, \bar\theta) &=& \phi_\mu (x) + \theta \;(s_{ab} \phi_\mu (x)) +
\bar\theta\; (s_b \phi_\mu (x)) + \theta \;\bar\theta\; (s_b s_{ab} \phi_\mu (x)),
\nonumber\\ \tilde {\cal F}^{(h)}_{\mu\nu} (x, \theta, \bar\theta) &=& C_{\mu\nu}
(x) + \theta (s_{ab} C_{\mu\nu} (x)) + \bar\theta (s_b C_{\mu\nu} (x))
+ \theta  \bar\theta (s_b s_{ab} C_{\mu\nu} (x)), \nonumber\\ \tilde
{\bar {\cal F}}^{(h)}_{\mu\nu} (x, \theta, \bar\theta) &=& \bar C_{\mu\nu} (x) +
\theta (s_{ab} \bar C_{\mu\nu} (x)) + \bar\theta (s_b \bar C_{\mu\nu} (x)) + 
\theta \bar\theta (s_b s_{ab} \bar C_{\mu\nu} (x)), \nonumber\\
\tilde {\cal F}^{(h)}_1 (x, \theta, \bar\theta) &=& C_1 (x) + \theta \;(s_{ab} C_1 (x))
+ \bar \theta\; (s_b C_1 (x)) + \theta \;\bar\theta\; (s_b s_{ab} C_1 (x)), \nonumber\\
\tilde {\bar {\cal F}}^{(h)}_1 (x, \theta, \bar\theta) &=& \bar C_1 (x) + \theta\; (s_{ab}
\bar C_1 (x))
+ \bar \theta\; (s_b \bar C_1 (x)) + \theta\;\bar\theta\; (s_b s_{ab} \bar C_1 (x)), \nonumber\\
\tilde {\cal F}^{(h)}_2 (x, \theta, \bar\theta) &=& C_2 (x) + \theta\;(s_{ab} C_2 (x))
+ \bar \theta\; (s_b C_2 (x)) + \theta\;\bar\theta\; (s_b s_{ab} C_2 (x)), \nonumber\\
\tilde {\bar {\cal F}}^{(h)}_2 (x, \theta, \bar\theta) &=& \bar C_2 (x) 
+ \theta\;(s_{ab} \bar C_2 (x))
+ \bar \theta\; (s_b \bar C_2 (x)) 
+ \theta\;\bar\theta\; (s_b s_{ab} \bar C_2 (x)), 
\end{eqnarray}
where the {\it proper} (anti-)BRST symmetry transformations are denoted by $s_{(a)b}$ and the superscript $(h)$, on
the superfields, stands for the super-expansions of these superfields obtained after the application of HC
(cf. (4)).

Furthermore, it can be checked that the anticommutativity 
property (i.e. $s_b s_{ab} + s_{ab} s_b = 0$) of $s_{(a)b}$
on the following basic  fields [8,9] 
\begin{eqnarray}
\{s_b, s_{ab} \} \;A_{\mu\nu\eta} = 0, \;\qquad 
\{s_b, s_{ab} \} \; C_{\mu\nu} = 0, \;\qquad
\{s_b, s_{ab} \}\; \bar C_{\mu\nu} = 0,
\end{eqnarray}
is true only when the Curci-Ferrari type restrictions (8) are satisfied. The property of the anticommutativity
of the (anti-)BRST symmetry transformations is trivially obeyed in the case of the rest of the fields of 
our present $D$-dimensional Abelian 3-form gauge theory. 
Finally, one can write down the coupled [but equivalent] (anti-)BRST invariant Lagrangian densities 
for the above Abelian 3-form gauge theory as (see, e.g. [9] for details)
\begin{eqnarray}
{\cal L}_B &=& {\displaystyle \frac{1}{24} H^{\mu\nu\eta\xi} H_{\mu\nu\eta\xi}
+ s_b s_{ab} \Bigl ( \frac{1}{2} \bar C_2 C_2 - \frac{1}{2} \bar C_1 C_1 + \frac{1}{2} \bar C_{\mu\nu} C^{\mu\nu}}
\nonumber\\
&-& \bar \beta^\mu \beta_\mu - {\displaystyle \frac{1}{2} \phi^\mu \phi_\mu - \frac{1}{6} B^{\mu\nu\eta} B_{\mu\nu\eta} \Bigr )}, 
\end{eqnarray}
\begin{eqnarray}
{\cal L}_{\bar B} &=& {\displaystyle \frac{1}{24} H^{\mu\nu\eta\xi} H_{\mu\nu\eta\xi}
- s_{ab} s_{b} \Bigl ( \frac{1}{2} \bar C_2 C_2 - \frac{1}{2} \bar C_1 C_1 + \frac{1}{2} \bar C_{\mu\nu} C^{\mu\nu}}
\nonumber\\
&-& \bar \beta^\mu \beta_\mu - {\displaystyle \frac{1}{2} \phi^\mu \phi_\mu 
- \frac{1}{6} B^{\mu\nu\eta} B_{\mu\nu\eta} \Bigr )}. 
\end{eqnarray}
The first Lagrangian density ${\cal L}_B$ is trivially invariant under the BRST transformations $s_b$. On the
other hand, the second Lagrangian density $ {\cal L}_{\bar B}$ is trivially invariant under the anti-BRST symmetry 
transformations $s_{ab}$. One can check that, under $s_{ab}$, the first Lagrangian density ${\cal L}_B$
transforms to a total
derivative plus terms that are zero on the constrained surface defined by the CF type restrictions (8). Precisely,
similar is the situation with the Lagrangian density ${\cal L}_{\bar B}$ under the nilpotent 
BRST transformations $s_b$. 
%Thus, ${\cal L}_B$ and $ {\cal L}_{\bar B}$ are equivalent on the constrained surface
%in $D$-dimensional spacetime manifold that is defined by the field equations (8).
   
\section{Comments on CF type restrictions}   

It is well-known that a gauge theory is always endowed with a local gauge symmetry that is generated by the first-class constraints 
in the language of Dirac's prescription for the classification scheme. Thus, the decisive
feature of a gauge theory is the existence of first-class constraints on the theory.  When any arbitrary $p$-form
gauge theory is discussed, within the framework of the BRST formalism, the above local gauge symmetry is traded with
the supersymmetric-type (anti-)BRST symmetries $s_{(a)b}$ which turn out to be nilpotent ($s_{(a)b}^2 = 0$) of
order two. Furthermore, the other sacrosanct feature of the latter symmetries is the absolute anticommutativity
(i.e. $s_b s_{ab} + s_{ab} s_b = 0$). The anticommutativity property is achieved only 
due to the presence of CF type restrictions.
Thus, the clinching feature of any arbitrary $p$-form gauge theory, within the framework of BRST formalism, is the
existence of the (anti-)BRST invariant CF type restrictions. For the Abelian 1-form gauge theory, the CF type
restriction is {\it trivial}. However, it is {\it non-trivial} for all the rest of the gauge theories.
Finally, the first-class constraints of the original gauge theory are encoded in the physicality criteria
$Q_b |phys> = 0$ where $Q_b$ is the conserved and nilpotent BRST charge. This condition, in BRST formalism,
enforces all the physical quantum states
to be annihilated by the operator form of the first-class constraints of the original theory.

\section{Conclusions}

In this presentation, it has been emphasized that the Bonora-Tonin's superfield approach [5,6] to BRST formalism
always leads to the derivation of the proper (i.e. off-shell nilpotent and absolutely anticommuting) (anti-)BRST
symmetry transformations for a given $p$-form gauge theory in any arbitrary $D$-dimensions of spacetime. Furthermore,
this geometrical superfield formalism [5,6] {\it necessarily} entails upon any arbitrary $D$-dimensional $p$-form gauge theory to be endowed with the (anti-)BRST invariant CF type restriction(s) which, ultimately, lead to the absolute anticommutativity of the (anti-)BRST symmetry transformations and the derivation of the coupled (but equivalent)
Lagrangian densities. It turns out that the CF condition, for the simple case of Abelian $U(1)$ 1-form gauge theory,  
is trivial. As a consequence, there is a single Lagrangian density for this theory that respects the (anti-)BRST
symmetries {\it together}. This is {\it not} the case, however, for even the non-Abelian $SU(N)$ 1-form gauge theory and all the rest of the (non-)Abelian higher $p$-form ($p \geq 2$) gauge theories in any arbitrary $D$-dimensions of sapcetime.\\
%Thus, the hallmark of a $p$-form gauge theory, within the framework of the BRST formalism, is the existence of (anti-)BRST invariant CF-type restriction(s).\\

\noindent
{\bf Acknowledgement}\\

\noindent
Travel support from DST, Govt. of India, is gratefully acknowledged. 
%Thanks are also due to organizers for their warm hospiatlity. 
%also due to A. Isaev, E. Ivanov and S. Krivonos for education in theoretical high energy physics
%at BLTP, JINR, Dubna, Moscow Region, Russia.  \\

\end{document}